\documentclass{article}

\usepackage{PRIMEarxiv}

\usepackage[utf8]{inputenc} 
\usepackage[T1]{fontenc}    
\usepackage{hyperref}       
\usepackage{url}            
\usepackage{booktabs}       
\usepackage{amsfonts}       
\usepackage{nicefrac}       
\usepackage{microtype}      
\usepackage{lipsum}
\usepackage{fancyhdr}       
\usepackage{graphicx}       
\usepackage{subcaption}
\usepackage{amsmath}
\usepackage{algorithm}
\usepackage{algpseudocode}
\graphicspath{{media/}}     

\title{Realtime Dynamic Gaze Target Tracking and Depth-Level Estimation
\thanks{\textit{\underline{Corresponding author}}: 
\textbf{Esmaeil Seraj <eseraj@ford.com>, Digital Vehicle Technology and AI, Ford Motor Company.}} 
}

\author{
  Esmaeil Seraj, Harsh Bhate, Walter Talamonti \\
  Digital Vehicle Technology and AI \\
  Ford Motor Company \\
  Dearborn, MI, USA \\
  \texttt{\{eseraj, hbhate, wtalamo1\}@ford.com} \\
}

\begin{document}
\maketitle

\begin{abstract}
The integration of Transparent Displays (TD) in various applications, such as Heads-Up Displays (HUDs) in vehicles, is a burgeoning field, poised to revolutionize user experiences. However, this innovation brings forth significant challenges in realtime human-device interaction, particularly in accurately identifying and tracking a user's gaze on dynamically changing TDs. In this paper, we present a two-fold robust and efficient systematic solution for realtime gaze monitoring, comprised of: (1) a tree-based algorithm for identifying and dynamically tracking gaze targets (i.e., moving, size-changing, and overlapping 2D content) projected on a transparent display, in realtime; (2) a multi-stream self-attention architecture to estimate the depth-level of human gaze from eye tracking data, to account for the display's transparency and preventing undesired interactions with the TD. We collected a real-world eye-tracking dataset to train and test our gaze monitoring system. We present extensive results and ablation studies, including inference experiments on System on Chip (SoC) evaluation boards, demonstrating our model's scalability, precision, and realtime feasibility in both static and dynamic contexts. Our solution marks a significant stride in enhancing next-generation user-device interaction and experience, setting a new benchmark for algorithmic gaze monitoring technology in dynamic transparent displays.
\end{abstract}


\section{Introduction}
\label{sec:intro}
Transparent Displays (TDs) are cutting-edge visual technologies that allow users to see digital content superimposed over physical environments with a variety of applications in dynamic Head-Up Displays (HUDs) in vehicles~\cite{riegler2019augmented, lindemann2018supporting, kim2009simulated}, augmented reality glasses~\cite{kim2013exploring, palinko2013towards, smith2020hit}, and smart windows in commercial buildings~\cite{mustafa2023smart}. Their ability to blend digital information with the real world offers significant advancements in fields such as navigation, interactive advertising, robotics~\cite{seraj2021hierarchical, seraj2021adaptive, seraj2022multi, seraj2020firecommander}, and immersive user interfaces and feedback~\cite{seraj2023enhancing, konan2021iterated, seraj2022embodied, seraj2023embodied, seraj2020coordinated}.
\begin{figure}[t]
  \centering
   \includegraphics[width=\linewidth]{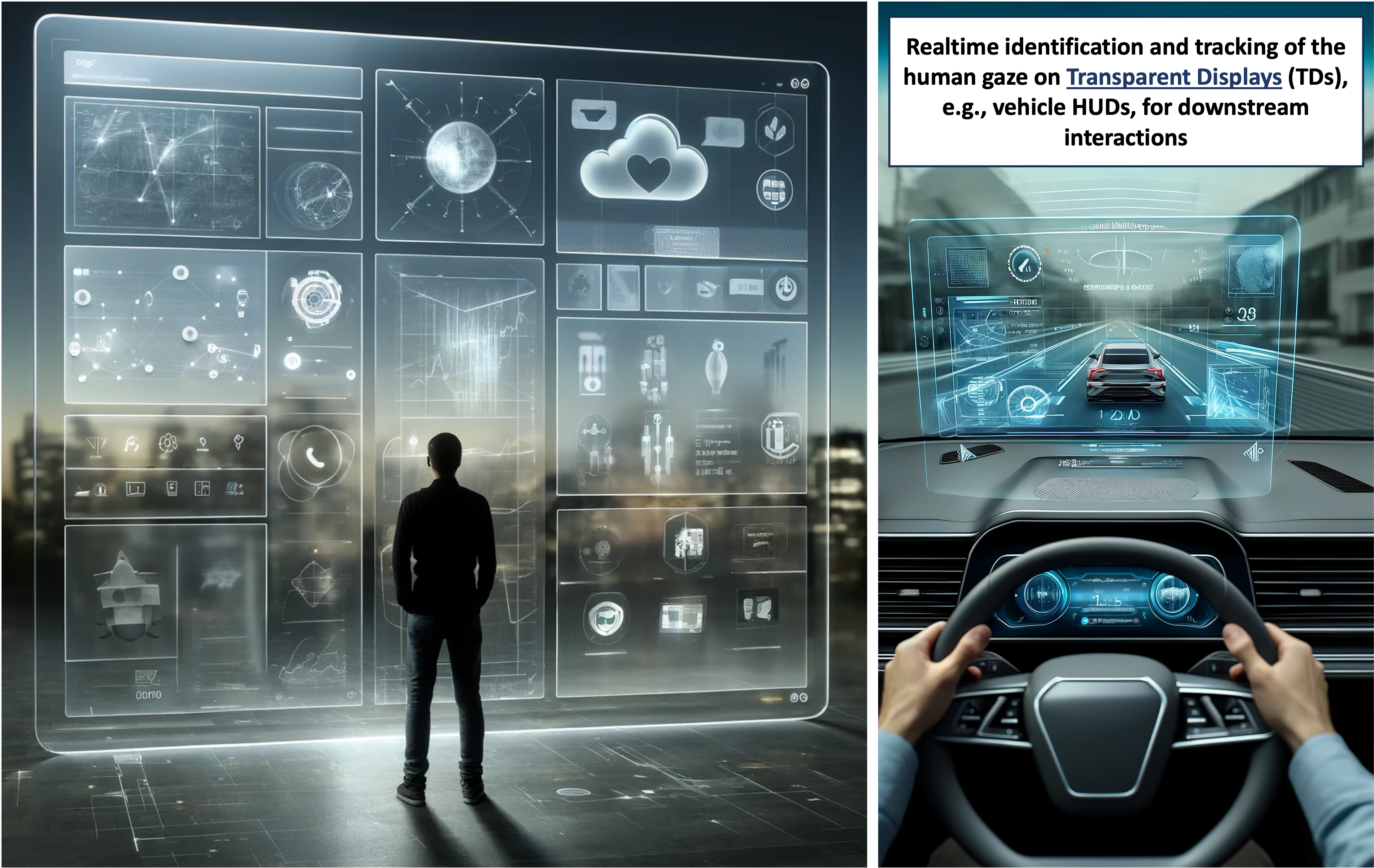}
   \caption{An example of a Transparent Display (TD), showing a large-scale dynamic Heads-Up Display (HUD) in a vehicle. A systematic solution is required to identify and smoothly track human gaze for downstream AR actions on projected HUD content, in realtime (image credit to DALL.E).}
   \label{fig:ar_windshield}
   \vspace*{-0.5cm}
\end{figure}

Imagine a transparent display, such as a dynamic HUD in a vehicle, that not only shows essential metrics like speed, fuel levels, and engine status but also overlays navigational cues directly onto the road ahead, highlighting paths, directions, pedestrians, and other vehicles~\cite{lindemann2018supporting, riegler2019augmented, jose2016comparative}. Beyond practical utilities, such dynamic HUDs could enhance the journey by identifying points of interest, e.g., service stations, or even serve as platforms for entertainment and work-related activities. However, realizing this vision introduces significant challenges, particularly in tracking the user's gaze across an ever-changing array of widgets and information layers projected onto the transparent display. Moreover, the accurate estimation of gaze depth levels is crucial, especially because of the display's transparency and the potential for the human gaze to interact with or pass through specific widgets, necessitating a system that can precisely discern the focus of a user's attention between virtual overlays and real-world objects to enhance both interactivity and safety~\cite{kwon20063d}. The dynamic nature of this problem, coupled with the need for real-time processing, sets a complex problem space for effectively identifying and monitoring what the user is focusing on at any given moment.

\begin{figure}[t]
  \centering
   \includegraphics[width=\linewidth]{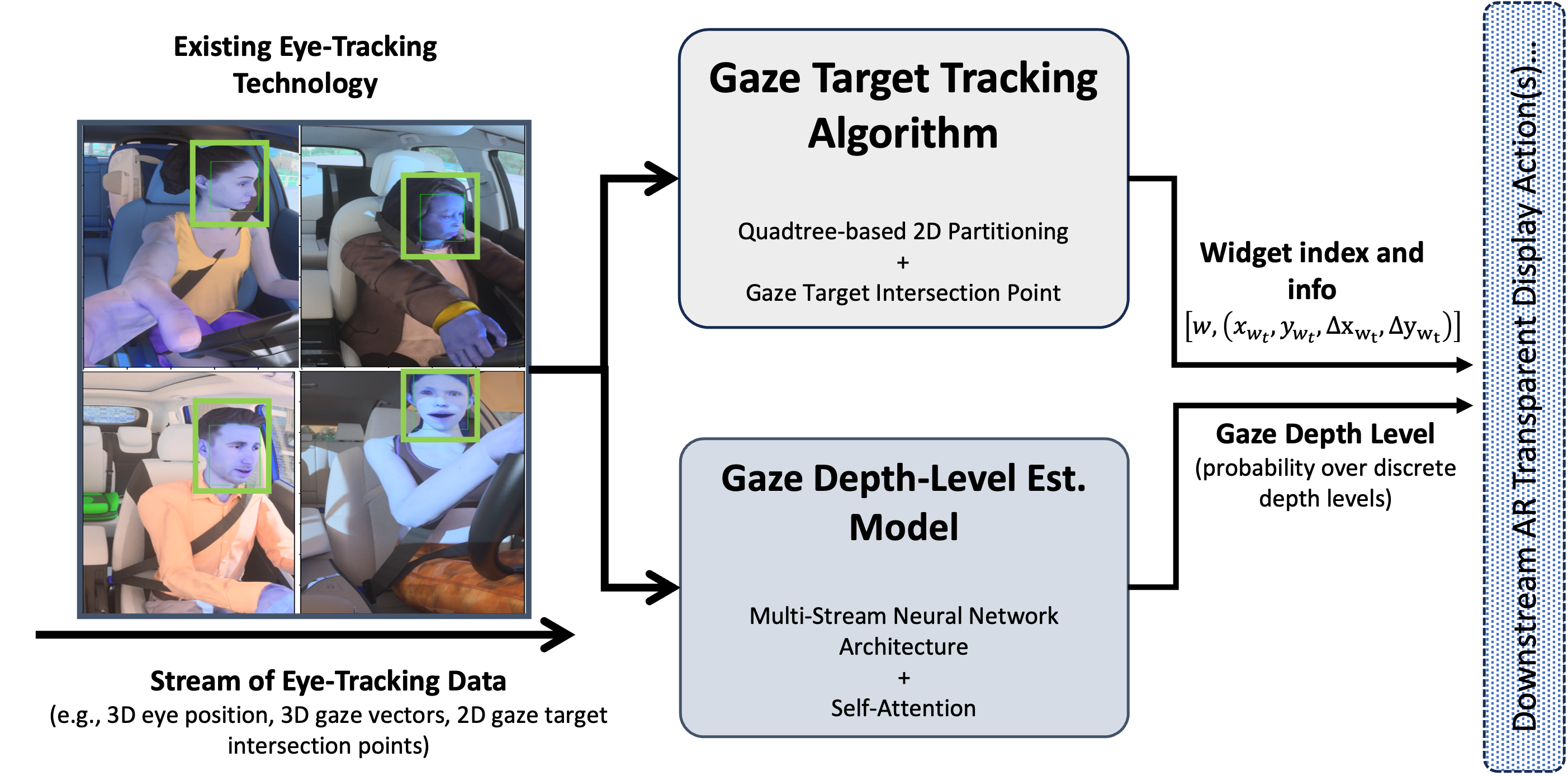}
   \caption{High-level overview of our gaze monitoring system. The gaze target tracking and gaze depth-level estimation modules work in parallel to robustly determine the focus of human gaze on the projected 2D widgets on the TD. Our system receives realtime eye-tracking data obtained from existing technology.}
   \label{fig:system_architecture}
   \vspace*{-0.5cm}
\end{figure}

In this work, we introduce a systematic approach to realtime and dynamic gaze monitoring in transparent displays (TDs). Our solution is comprised of two modules: (1) a tree-based gaze target tracking, and (2) a multi-stream neural network architecture for gaze depth-level estimation. The first module includes a robust Quadtree spatial partitioning framework, designed to efficiently manage and query the spatial distribution of display widgets. This framework enables our algorithm to quickly and accurately determine the user's gazed target among multiple overlapping widgets, adapting seamlessly to changes in widget positions and sizes. The second module in our system is a multi-stream attention model for enhanced human gaze depth-level estimation from composite and multi-dimensional eye tracking data. This model works in parallel (see Fig.~\ref{fig:system_architecture}) with the first module to accurately determine the focus of a user's attention between virtual widgets and real-world objects behind the TD. Our evaluations confirmed the accuracy, scalability, and realtime feasibility of our solution. \textbf{Our key contributions are as follows:}
\begin{enumerate}
    \item We propose a new tree-based algorithm with realtime feasibility and high precision for identifying and tracking dynamic (i.e., moving, overlapping) gaze targets projected as 2D widgets on displays, such as TDs.

    \item We propose a light-weight multi-stream self-attention model to accurately estimate human gaze depth-level via composite, multi-dimensional eye tracking data.

    \item We collect real-world human data to train and test our model feasibility, demonstrating the robustness of our proposed architecture for gaze depth prediction.

    \item We present extensive results, including inference performance on representative hardware evaluation boards, confirming realtime deployability of our solutions.
\end{enumerate}


\section{Related Work}
\label{sec:Related Work}
Recent advancements in AR technologies has garnered significant attention, especially within the automotive industry~\cite{fathiazar2023using, shahid2013eye, kang2021real}. In this section, we overview the critical strands of research that inform and contextualize our study. 

\textbf{Transparent Displays (TDs) in Vehicles} --  AR screens and displays have emerged as a frontier technology in enhancing driving experience and safety~\cite{riegler2019augmented, lindemann2018supporting, kim2009simulated, natarajan2023human}. Initial studies and implementations have focused on overlaying navigation data~\cite{palinko2013towards, jose2016comparative}, safety alerts~\cite{kim2013exploring, smith2020hit}, and contextual information~\cite{lindemann2018supporting, riegler2019augmented, kim2009simulated} directly onto the driver's field of view. Although these preliminary solutions have paved the way towards realizing TDs in vehicles, they have focused predominantly on static overlays and often lack adaptability to realtime dynamic gaze target tracking in dense scenarios with overlapping and moving targets.

\textbf{Gaze Target Tracking} -- Tracking gazed targets~\cite{beymer2003eye, fang2021dual, seraj2024invehicle} is pivotal in human-computer interaction for understanding user engagement and focus, particularly in environments enriched with AR content~\cite{majaranta2014eye, drewes2010eye}. In the application of dynamic HUDs in vehicles, tracking gazed targets (e.g., projected 2D widgets) of a driver presents unique challenges due to the high-speed  and dynamic nature of driving (e.g., moving vehicles and pedestrians) and the need for instantaneous feedback. Research in this domain has explored various eye-tracking technologies~\cite{kang2021real, amarnag2003real, chau2005real, xu2018real} to determine driver focus areas on the windshield~\cite{fathiazar2023using, shahid2013eye, kang2021real}. While promising, these methods often struggle with latency issues and inaccuracies in rapidly changing visual scenes, leading to unreliable gaze monitoring. Moreover, these prior work do not explicitly account for the gaze depth, which is critical in transparent displays to avoid undesired interactions. Our solution is tailored for the dynamic setting of TDs and leverages an advanced gaze tracking framework optimized for low latency and high accuracy, while simultaneously providing an estimation of the gaze depth.

\textbf{Gaze Depth Estimation} -- 3D Gaze depth estimation, particularly in the context of 2D interactions with \emph{transparent} displays, is critical, due to the need to discern whether a user's gaze is focused on virtual widgets or is merely passing through them. This distinction is critical to avoid undesired interactions with projected widgets, which could potentially distract the user or lead to misinterpretations of user intent. Prior work in gaze depth estimation spans classical model-based techniques~\cite{oney2020evaluation, kwon20063d, sun2015real, kuo2018depth, duchowski2014comparing, pfeiffer2012measuring} that often rely on geometric models of the eye and its movements, to more advanced data-driven approaches~\cite{lee2017estimating, villavicencio2023estimating, lee2017gaze, weier2018predicting} leveraging machine learning to predict gaze depth from eye-tracking data. Despite advancements, existing methods face challenges in terms of robustness, often being sensitive to slight variations in user gaze behavior~\cite{oney2020evaluation, duchowski2014comparing}, and performance~\cite{villavicencio2023estimating}. Our approach seeks to address these shortcomings by introducing a sophisticated multi-stream model equipped with intra- and inter-stream self-attention mechanisms to robustly capture and learn physical dependencies embedded within eye tracking data samples.

\section{Preliminaries}
\label{sec:Background}
Here, we present a rigorous formulation of our gaze monitoring problem and provide the requisite overview of the fundamental concepts in our algorithm design, including the Quadtree data structures and the self-attention mechanism.

\subsection{Problem Formulation}
\label{subsec:Problem Formulation}
One of the core challenges in augmenting transparent displays with AR content lies in accurately determining where on the screen the user is looking at any given moment, and identifying the specific 2D widget projected on the TD that is being gazed at. Given the dynamic nature of the widgets in our application, we need to track the gazed widget in realtime. As shown in Fig.~\ref{fig:system_architecture}, our framework is built on available realtime eye tracking data~\cite{kang2021real}. We utilize a commercially available eye tracking consumer camera, such as the SmartEye eye tracking camera, that can provide realtime streams of 3D eye positions, $e_i$, 3D eye rotations (gaze vectors for each eye), $r_i$, and 2D intersection points on a virtual plane in front of the user, $g_i$. 

Let $\mathcal{G} = \{(g_x, g_y) \mid g_x, g_y \in \mathbb{R}\}$ denote the continuous stream of gaze vector intersection points on the virtual plane (i.e., TD in our case), where each $g_i$ represents the normalized coordinates within the display surface. Consider a set of dynamic AR widgets $\mathcal{W} = \{w_1, w_2, \ldots, w_n\}$, available via the vehicle digital interfaces at any given time. Each widget $w_i$ is represented by a 4-tuple: $(x_i, y_i, \Delta x_i, \Delta y_i)$, where the $(x_i, y_i)$ is the top-left corner of a widget, and $\Delta x_i$ and $\Delta y_i$ are its width and height. The objective of the problem is to identify the widget $w_j \in \mathcal{W}$ that the gaze intersection point $g_i$ is targeting at any instant $t$. This problem involves realtime processing of the gaze stream $\mathcal{G}$ and dynamic updates to the set $\mathcal{W}$, reflecting changes in the AR content displayed on the (transparent) display. 

Additionally, in the context of 2D interactions with transparent displays, we also need to estimate the 3D gaze depth-level, to discern whether a user's gaze is focused on virtual 2D widgets or is merely passing through them. This distinction is critical to avoid undesired interactions with projected widgets. To this end, we propose a data-driven solution to tackle the notorious sensitivity issues of the classical gaze depth models~\cite{oney2020evaluation, duchowski2014comparing, lee2017estimating, villavicencio2023estimating}. Consider a dataset $\mathcal{D} = \{(s_i, l_i)\}_{i=1}^{\mathcal{N}}$ of $\mathcal{N}$ samples, where each sample $s_i$ is represented by a tuple of eye-tracking data and a label $l_i \in \{1, 2, \ldots, \mathcal{C}\}$ indicating the gaze depth-level categories. Each sample $s_i$ consists of gaze direction vectors for the left and right eyes, $r_i = ((x_{1i}, y_{1i}, z_{1i}), (x_{2i}, y_{2i}, z_{2i}))$, eye position vectors in space for the left and right eyes, $e_i = ((ex_{1i}, ey_{1i}, ez_{1i}), (ex_{2i}, ey_{2i}, ez_{2i}))$, and the 2D gaze intersection points. We consider three gaze depth-level categories: \{'\texttt{on-plane}', '\texttt{out-plane-near}', '\texttt{out-plane-far}'\}. The objective is then to learn a model to accurately predict the gaze depth-level by capturing the physical dependencies in a given input tuple of eye-tracking data, such as relations between eye positions, rotations (gaze vectors), and target intersection points. 

\begin{figure*}[t]
  \centering
   \includegraphics[width=\linewidth]{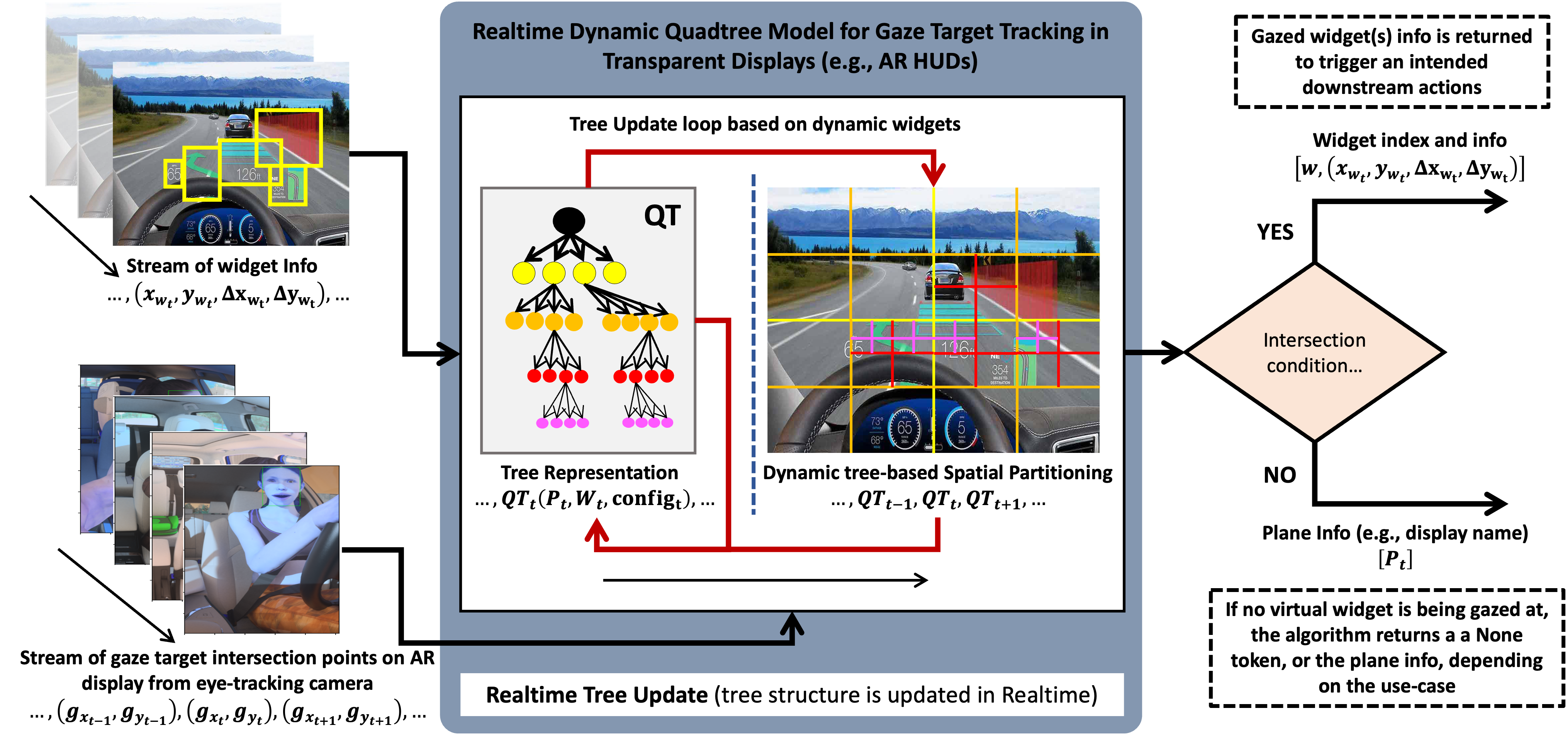}
   \caption{The architecture of the proposed Quadtree-based realtime dynamic gaze target tracking algorithm (i.e., the first module in our gaze monitoring system shown in Fig.~\ref{fig:system_architecture}) for transparent displays (e.g., here a dynamic HUD). At any given time, the algorithm receives the stream of the gaze target intersection points on the AR display, given by the eye-tracking camera, as well as the stream of widget information tuples. The tree structure is updated in realtime based on the received set of widget information and an efficient tree-traversal (i.e., Depth-First Search (DFS)) is performed to associate a gaze target intersection point to a leaf-node in the constructed tree. }
   \label{fig:quadtree_gaze_target_traking}
   \vspace*{-0.5cm}
\end{figure*}

\subsection{Quadtree Data Structures}
\label{subsec:Quadtree Data Structures}
A Quadtree is a tree data structure designed for partitioning a 2D space by recursively subdividing it into four quadrants or nodes. Mathematically, a Quadtree can be represented as $QT = (q, l)$, where $q$ is a node that can either be a leaf node containing data points or an internal node with four children (quadrants); and $l$ denotes the list of data points or objects within the leaf nodes. Each node in a Quadtree corresponds to a specific rectangular region of the space, with the root node encompassing the entire area. As a versatile data structure, Quadtrees find applications across various fields, including computer graphics~\cite{brainerd2016efficient}, image processing~\cite{gillespie1981tree}, and geographical information systems~\cite{samet1984geographic}, due to their efficiency in handling complex spatial queries and operations. For dynamic gaze target tracking in AR windshields, the Quadtree structure allows for rapid determination of which widgets are under the driver's gaze and efficient updates as widgets move, appear, or disappear.


\subsection{Self-Attention Mechanism}
\label{subsec:Self-Attention Mechanism}
The self-attention mechanism is a pivotal component in modern ML~\cite{shaw2018self, vaswani2017attention, seraj2024mixed, konan2023contrastive}. It enables a model to dynamically focus on different parts of the input sequence when processing a specific element of that sequence, thereby capturing contextual relationships regardless of their positional distance within the input. Given an input sequence $\mathbf{X} = [\mathbf{x}_1, \mathbf{x}_2, \ldots, \mathbf{x}_n]$ where each $\mathbf{x}_i \in \mathbb{R}^d$ represents a vectorized representation of the $i^{th}$ element in the sequence, the self-attention process transforms $\mathbf{X}$ into an output sequence $\mathbf{Z} = [\mathbf{z}_1, \mathbf{z}_2, \ldots, \mathbf{z}_n]$ where each $\mathbf{z}_i$ is computed as a weighted sum of linearly transformed input vectors. For each element in the sequence, the self-attention mechanism first computes three vectors through linear transformations of the input vector $\mathbf{x}_i$: a query vector, $\mathbf{q}_i = \mathbf{W}_Q \mathbf{x}_i$, a key vector, $\mathbf{k}_i = \mathbf{W}_K \mathbf{x}_i$, and a value vector, $\mathbf{v}_i = \mathbf{W}_V \mathbf{x}_i$, where $\mathbf{W}_Q$, $\mathbf{W}_K$, and $\mathbf{W}_V$ are the weight matrices for the query, key, and value vectors, respectively.

Accordingly, the attention score $a_{ij}$, representing the importance of the $j^{th}$ element's value vector $\mathbf{v}_j$ for computing the output $\mathbf{z}_i$, is calculated by taking the dot product of the query vector $\mathbf{q}_i$ with the key vector $\mathbf{k}_j$, given by $a_{ij} = \frac{\mathbf{q}_i^\top \mathbf{k}_j}{\sqrt{d_k}}$, where $d_k$ is the dimensionality of the key vectors, and the division by $\sqrt{d_k}$ is a scaling factor introduced to mitigate the vanishing gradients problem in Softmax. The attention scores for a particular query are then normalized across all keys using the Softmax function to obtain the attention weights, $\alpha_{ij} = \frac{\exp(a_{ij})}{\sum_{l=1}^{n} \exp(a_{il})}$. Finally, the output vector is then computed by $\mathbf{z}_i = \sum_{j=1}^{n} \alpha_{ij} \mathbf{v}_j$.








\section{Method}
\label{sec:Method}
Here, we delineate the high-level overview of our gaze monitoring system architecture and operational dynamics of our proposed solution. Using dynamic HUDs in vehicles as a running example, we then elaborate the two parallel modules in our framework for realtime dynamic gaze tracking and gaze depth-level estimation within the context of TDs.

\subsection{High-level Overview}
\label{subsec:System Architecture Overview}
The high-level schematic of our driver gaze monitoring framework for TDs is shown in Fig.~\ref{fig:system_architecture} for a dynamic HUD application. The gaze target tracking and gaze depth-level estimation modules work in parallel to robustly determine the focus of driver gaze on the virtual 2D widgets projected on the AR screen (e.g., HUD). Our system receives realtime eye-tracking data obtained from existing technology. 

As described in our problem formulation in Section~\ref{subsec:Problem Formulation}, the gaze target tracking module encompasses an efficient tree-based algorithm, the objective of which is to identify the widget $w_j \in \mathcal{W}$ that the user's gaze intersection point $g_i$ on the display is targeting at any instant $t$. Internally, this process includes three primary components: (1) the eye tracking module, for which we leverage existing technology capable of providing realtime eye and gaze tracking information; (2) the AR content management module, which is available in the device's onboard technology and provides realtime stream of dynamic widget(s) information as a set of 4-tuples, described in Section~\ref{subsec:Problem Formulation}; and (3) the dynamic Quadtree-based realtime gaze target tracking algorithm, which receives the information from parts (1) and (2), and performs an efficient tree traversal to return the desired widget information. Note that, in automotive applications, existing work on pedestrian and vehicle tracking for autonomous vehicles~\cite{ahmed2019pedestrian, camara2020pedestrian} can be leveraged to project predicted content on the vehicle display, which then can be fed into our algorithm to track driver gaze and improve situational awareness via notifications, in case of poor visibility.

The gaze depth-level estimation model, shown in Fig.~\ref{fig:system_architecture}, works in parallel with the gaze target tracking module to predict the probability of a user's gaze being focused on the widgets projected at some Virtual Image Distance (VID), or merely passing through the transparent display. This is to avoid undesired gaze-based downstream actions and is achieved via a new multi-stream attention model that provides this probability, given complex, multi-dimensional eye tracking data samples. Our proposed architecture is light-weight, including a few fully-connected layers and efficient vector-based self-attention operations which can be executed in realtime alongside the first module at inference.

\subsection{Dynamic Quadtree for Gaze Target Tracking}
\label{subsec:Dynamic Quadtree for Gaze Target Tracking}
The cornerstone of our method is the dynamic QT structure, which facilitates efficient management and querying of spatial data. Fig.~\ref{fig:quadtree_gaze_target_traking} and Alg.~\ref{alg:system_architecture} show the framework architecture and pseudocode, respectively, for our dynamic QT-based solution for realtime tracking of the gaze targets for TDs. A node in the Quadtree, $q$, can either be a leaf node containing a some widgets, $\{w_1, w_2, \cdots, w_k\}$, that fall within its quadrant or a parent node with four children, each representing one of the four subdivisions of $q$'s region. The subdivision process (line 17 in Alg.~\ref{alg:system_architecture}) continues until either of the following two conditions are met: (1) a leaf node contains no more than a predetermined number of widgets (i.e., we chose one in our setup), or (2) a minimum size of the quadrant is reached, ensuring efficient spatial querying and updating. Note that, only nodes that contain widgets are subdivided further into quadrant. Additionally, for all QT operations (i.e., adding, removing, moving, or finding a widget), an efficient tree traversal algorithm such as Breadth-First-Search (BFS) is leveraged (lines 12 and 17 in Alg.~\ref{alg:system_architecture}), allowing for rapid determination of which widgets are under the user's gaze and efficient updates as widgets move, appear, or disappear. This way, in case of overlapping widgets, the algorithm returns the information of all widgets inside a leaf node.
\begin{algorithm}
\caption{Dynamic QT for Gaze Target Tracking}\label{alg:system_architecture}
\begin{algorithmic}[1]
\State \textbf{Initialize} eye tracking, AR content management, $QT$, max. num. of widgets in a node, $T$, min. node size, $L$
\Function{MainLoop}{}
    \While{system is active}
        \State $G_t \gets$ Get current gaze intersection points
        \State $\mathcal{W}_t \gets$ Get current set of widgets
        \State \Call{ProcessGaze}{$G_t$}
        \State \Call{UpdateQuadtree}{$\mathcal{W}_t$}
    \EndWhile
\EndFunction
\Function{ProcessGaze}{$G_t$}
    \For{each $g_i$ in $G_t$}
        \State $w_j \gets$  Traverse $QT$ to find current gaze target
        \State Return $w_j$ if $w_j$, else return $None$
    \EndFor
\EndFunction
\Function{UpdateQuadtree}{$\mathcal{W}_t$, $T$, $L$}
    \State Rebuild $QT$ given the new $\mathcal{W}_t$, based on $T$ \& $L$
\EndFunction
\end{algorithmic}
\end{algorithm}
\vspace*{-0.5cm}

\subsubsection{Quadtree Initialization and Widget Insertion}
\label{subsubsec:Quadtree Initialization and Widget Insertion}
The QT is initialized to encompass the entire TD area, represented as a normalized 2D space $[0,1] \times [0,1]$. Each widget $w_i \in \mathcal{W}$ is inserted into the QT based on its spatial coordinates and dimensions. In the insertion process, if a node exceeds the maximum capacity $T$, it subdivides into four quadrants. The widget is inserted into the appropriate quadrant(s) through BFS tree traversal. If a widget spans multiple quadrants, it is referenced in each relevant quadrant to ensure accurate gaze tracking.

\subsubsection{Widget Identification and Dynamic Updates}
\label{subsubsec:Gaze Point Processing and Widget Identification}
For every gaze intersection point $g_t = (g_x, g_y)$ at time $t$, the system traverses the QT from the root, narrowing down to the quadrant(s) that contain the gaze point and collects all widgets within the leaf node that intersect with $g_t$. The system then returns the topmost widget(s) as the gaze target based on predefined criteria, such as widget priority or z-order. Dynamic updates—widget movements, additions, or removals—are handled in realtime, given the set of all widgets, $\mathcal{W}_t$ at each timestep, to reflect changes in the AR display. This dynamic update capability ensures that the QT accurately represents the current state of the TD.

\begin{figure*}[t]
  \centering
   \includegraphics[width=\linewidth]{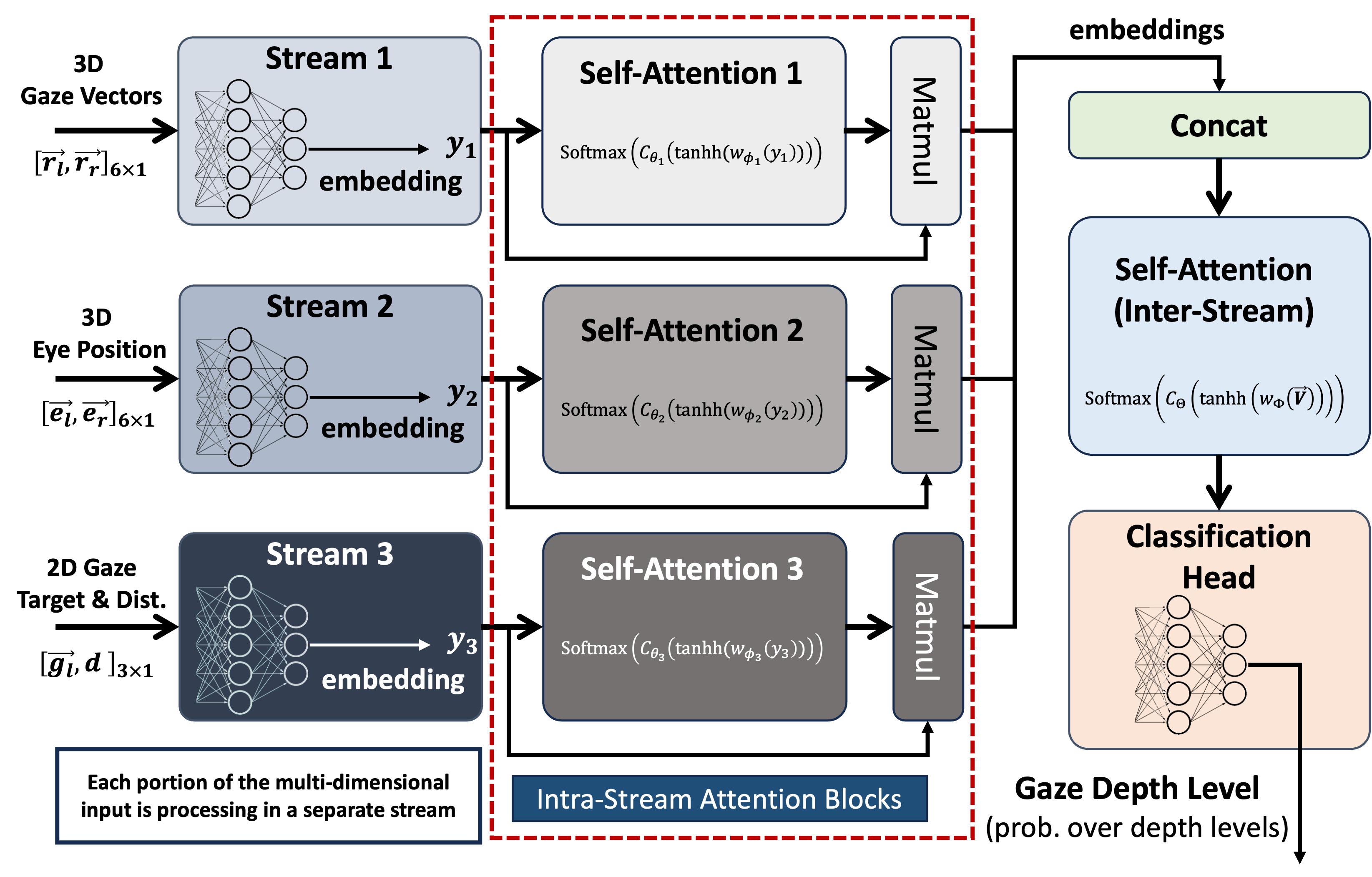}
   \caption{The proposed multi-stream attention model for gaze depth-level estimation from composite multi-dimensional eye-tracking data. To capture the physical dependencies within and between different portions of the eye-tracking data, we leverage intra- and inter-stream self-attention layers. The intra-stream attention layers capture the physical relations between the elements of each data section (i.e., connection between left and right eye or their internal X, Y, and Z values), while the inter-stream attention captures the relation between eye position, eye rotation (gaze vectors), and 2D gaze target intersection points. The output is a probability over discrete gaze depth levels.}
   \label{fig:gaze_depth_model}
   \vspace*{-0.5cm}
\end{figure*}

\subsection{Multi-Stream Attention Model for Gaze Depth-Level Estimation from Eye-Tracking Data}
\label{subsec:Multi-Stream Attention Model for Gaze Depth-Level Estimation}
Prior work have proposed and investigated a mathematical model for estimating the 3D depth of human gaze from eye tracking data, such as eye positions and rotations (i.e., gaze direction vectors), shown in Eq.~\ref{eq:classical_gaze_depth_model}. This model uses the concept of triangulation from geometry~\cite{kuo2018depth}, in which $\Delta e_x$ is the inter-ocular distance, $d_v$ is the distance to a virtual plane in front of the human (e.g., a monitor screen), and $\Delta g_x = |g_{x_l} - g_{x_r}|$ is the horizontal distance between the intersection points of the gaze vectors from left and right eyes on the virtual plane~\cite{kuo2018depth}. Nevertheless, this model is reportedly sensitive to minor changes in its parameters and typically does not meet performance requirements~\cite{duchowski2014comparing, kuo2018depth, lee2017estimating}.
\begin{equation}
    \label{eq:classical_gaze_depth_model}
    g_{\text{depth}} = \frac{\Delta e_x d_v}{\Delta e_x - \Delta g_x}
\end{equation}

To achieve a robust estimating of the gaze depth given eye tracking data, we instead propose approximating the model in Eq.~\ref{eq:classical_gaze_depth_model} by introducing a sophisticated model that leverages the power of attention mechanisms within a multi-stream neural network framework. Our motivation is to build a model that can work with composite, multi-dimensional eye tracking data format, capable of accurately categorizing the nature and depth of human gaze based on this data. This essentially ensures that the necessary variables to approximate a 3D gaze depth model based on Eq.~\ref{eq:classical_gaze_depth_model} are leveraged and their physical dependencies, such as relations between eye positions, rotations (gaze vectors), and target intersection points, are accounted for.

To achieve this, we propose a multi-stream attentional model architecture (see Section~\ref{subsubsec:Gaze Depth Model Architecture} for details) that is trained on collected eye tracking data. The data for each gaze instance includes gaze direction vectors for each eye, $\left[\Vec{r}_l, \Vec{r}_r\right]_{6\times 1}$, eye position coordinates, $\left[\Vec{e}_l, \Vec{e}_r\right]_{6\times 1}$, target points on a screen, and the distance from the eyes to the screen, $\left[\Vec{g}_t, d_v\right]_{3\times 1}$. Each data sample is labeled into three categories, \{'\texttt{on-plane}', '\texttt{out-plane-near}', '\texttt{out-plane-far}'\}, (see Section~\ref{subsec:Data Collection} for more details regarding our data collection process).

\subsubsection{Gaze Depth Model Architecture}
\label{subsubsec:Gaze Depth Model Architecture}
Our proposed model employs a multi-stream neural network architecture with intra- and inter-stream attention mechanisms. Fig.~\ref{fig:gaze_depth_model} shows the proposed model for gaze depth-level estimation from composite multi-dimensional eye-tracking data. To capture the physical dependencies within and between different segments of the eye-tracking data, we leverage intra- and inter-stream self-attention layers. Each stream operates by generating a local embedding, $\mathbf{y}_j$, from their respective input data, which are then passed through a stream-specific self-attention mechanism, $\mathbf{a}_j = \alpha_j(\mathbf{y}_j) = \text{softmax}\left( \mathbf{W}_{c_j}^T \tanh(\mathbf{W}_{a_j} \mathbf{y}_j + \mathbf{b}_{a_j}) \right) \odot \mathbf{y}_j$. Here, $\mathbf{W}_{a_j}$ and $\mathbf{W}_{c_j}$ are the weights of the attention mechanism for stream $j$, and $\odot$ denotes element-wise multiplication. These intra-stream attention layers capture the physical relations between the elements of each data section (i.e., connection between left and right eye or their internal $X$, $Y$, and $Z$ values). The output embeddings of each stream, $a_j$, are then concatenated, $\Vec{\mathbf{a}} = \left[\mathbf{a}_1; \mathbf{a}_2; \mathbf{a}_3\right]$, and fed into the inter-stream self-attention layer, $\mathbf{a} = \beta(\Vec{\mathbf{a}}) = \text{softmax}\left( \mathbf{W}_c^T \tanh(\mathbf{W}_a \Vec{\mathbf{a}} + \mathbf{b}_a) \right) \odot \Vec{\mathbf{a}}$. The inter-stream attention captures the high-level relation between eye position, gaze vectors, and 2D gaze intersection points. The final gaze depth category prediction is made by passing the output of the inter-stream attention through a classification head which predicts a probability distribution over gaze depth categories for a given sample.

Note that, the input to our multi-stream attention model are small numerical vectors and with only a few fully-connected and attention mechanisms, our architecture can in fact perform fast feedforward operations at inference, making the realtime process feasible. Alg.~\ref{alg:depth_model_training} presents the process of training our proposed multi-stream attention model architecture for gaze depth-level classification.
\begin{algorithm}
\caption{Training the Multi-Stream Gaze Depth Model}\label{alg:depth_model_training}
\begin{algorithmic}[1]
\State \textbf{Input:} Dataset $\mathcal{D} = \{(s_i, l_i)\}_{i=1}^{N}$, $lr=\eta$, epochs $E$
\For{epoch $= 1$ to $E$}
    \For{each batch $(\mathbf{S}, \mathbf{Y})$ in $\mathcal{D}$}
        \State $\mathbf{S}_{\text{stream}} \gets$ Split $\mathbf{S}$ into stream-specific inputs
        \State $\mathbf{H}_{\text{stream}} \gets$ Apply stream networks to $\mathbf{S}_{\text{stream}}$
        \State $\mathbf{A}_{\text{stream}} \gets$ Apply intra-stream att. to $\mathbf{H}_{\text{stream}}$
        \State $\mathbf{A}_{\text{combined}} \gets$ Combine $\mathbf{A}$ across streams
        \State $\mathbf{A} \gets$ Apply inter-stream attention to $\mathbf{A}_{\text{combined}}$
        \State $\hat{\mathbf{Y}} \gets$ Apply final classification layer to $\mathbf{A}$
        \State $\mathcal{L} \gets$ Compute loss between $\hat{\mathbf{Y}}$ and $\mathbf{Y}$
        \State Backprop and update model
    \EndFor
\EndFor
\end{algorithmic}
\end{algorithm}

\begin{figure*}[t]
  \centering
   \includegraphics[width=\linewidth]{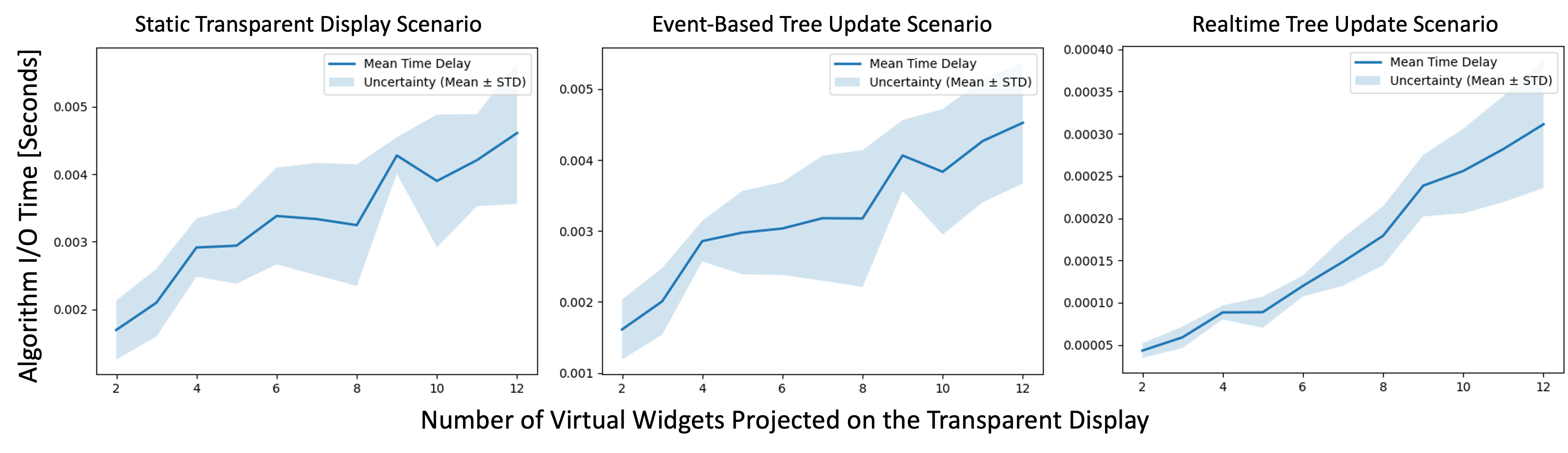}
   \caption{Scalability analysis for our tree-based target tracking algorithm. Results demonstrate system's input-output latency (mean$\pm$STD) in three experimental scenarios described in Section~\ref{subsec:Results and Discussion}, over a range of virtual widgets projected on a simulated transparent display. As shown, the framework can approximately linearly scale to the number of widgets, demonstrating a realtime-feasible implementation.
}
   \label{fig:latency_results}
   \vspace*{-0.5cm}
\end{figure*}

\section{Evaluation}
\label{sec:Evaluation}
The efficacy of our proposed real-time dynamic gaze target identification and tracking system for (transparent) displays is rigorously evaluated through a series of experiments designed to simulate varying conditions. We also trained and evaluated our proposed gaze depth-level estimation model on a set of eye tracking data collected from live participants. Emphasis is placed on evaluating accuracy, latency, and update frequency, crucial metrics that underscore the system's viability for practical realtime deployment.

\subsection{Eye Tracking Data Collection}
\label{subsec:Data Collection}
We leveraged existing eye-tracking technology (e.g., the SmartEye camera) for collecting real eye-tracking data in a vehicle HUD case-study. The camera was mounted on top of the center stack facing the cabin in a stationary vehicle. We recruited six subjects collecting their eye-tracking data in various scenarios. For the gaze depth scenario, we chose a simple setup by discretizing the depth into three levels, on-plane, out-plane-near, and out-plane-far, which correspond to the participant's gaze focused on a sticker on the windshield, on an object (e.g., orange cone) placed outside and near the vehicle (i.e., $\sim 6$ meters), and on an object placed outside and far from the vehicle (i.e., $\sim 12$ meters). We collected gaze depth data during $30$-seconds iterative sessions per subject, creating a dataset of $\sim 11,000$ samples, using our eye tracking camera operating under $60$Hz frequency.

\subsection{Gaze Target Tracking Algorithm Performance}
\label{subsec:Results and Discussion}
We experimented and evaluated our dynamic gaze target tracking algorithm for (transparent) displays in three different simulated scenarios: (1) static TD (no tree updates), (2) event-based tree update (i.e., QT is only updated when a change is made to the set of widgets), and (3) realtime tree update, where the tree is constantly updated at each step to account for dynamic widgets. Our algorithm demonstrated high precision performance, achieving $100\%$ simulation accuracy, in all three scenarios, including the realtime tracking of gaze targets with highly dynamic widgets. Please see the attached supplementary video for a demonstration of the algorithm performance in simulation.

\textbf{Scalability Analysis} -- We analyzed the scalability of our tree-based target tracking algorithm to evaluate its feasibility across increased number of widgets. Results in Fig.~\ref{fig:latency_results}, demonstrate system's input-output latency (mean$\pm$STD) over a range of virtual widgets projected on the simulated TDs. For cases (1) and (2) in aforementioned experimental scenarios, each point on the plots in Fig.~\ref{fig:latency_results} demonstrates the average I/O latency during $30$ seconds of algorithm operation, while for case (3), values show the average latency during $1$ second of operation. Presented results are achieved on a laptop equipped with 12th Gen Intel Core i7-12850HX CPU (4.80 GHz). As shown, the framework can approximately linearly scale to the number of widgets, achieving an average of $0.31$ miliseconds latency over $60$ samples in the aggressive realtime scenario with 12 moving widgets, demonstrating a realtime-feasible implementation.

\begin{figure}
  \centering
  \begin{subfigure}{0.49\linewidth}
    \includegraphics[width=\linewidth]{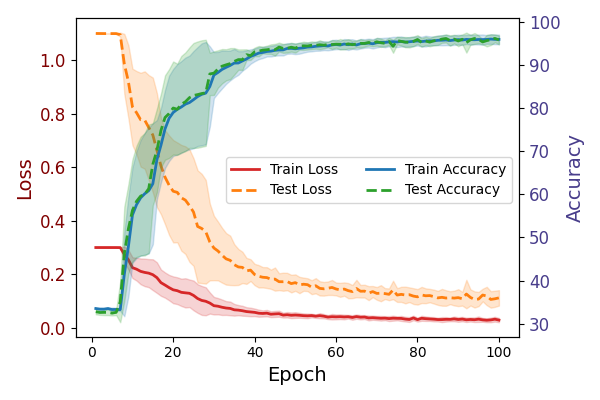}
    \caption{Full model performance (loss and accuracy).}
    \label{fig:depth_model_results}
  \end{subfigure}
  \hfill
  \begin{subfigure}{0.49\linewidth}
    \includegraphics[width=\linewidth]{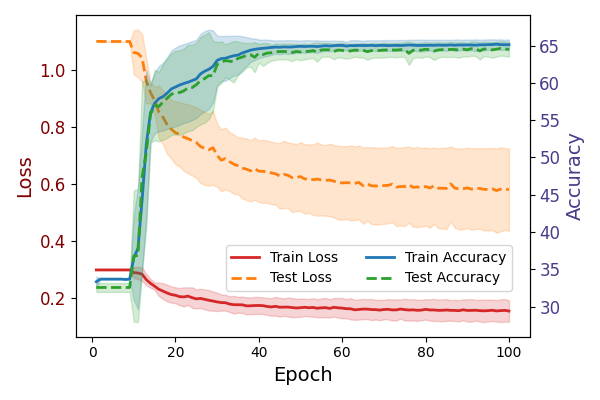}
    \caption{Model performance w/o the intra-steam attention layers.}
    \label{fig:depth_model_ablation_results}
  \end{subfigure}
  \caption{Gaze depth-level estimation model performance; the full architecture (Fig.~\ref{fig:depth_model_results}) vs. the model without the intra-steam self-attention layers (Fig.~\ref{fig:depth_model_ablation_results}). Results show that the incorporated self-attention layers can effectively capture the physical dependencies embedded in our recorded real-world eye tracking data.}
  \label{fig:short}
  \vspace*{-0.5cm}
\end{figure}

\subsection{Gaze Depth-Level Model Performance}
\label{subsec:Results and Discussion - Depth Model}
We trained, evaluated, and tested our multi-stream attention model for 3D gaze depth-level estimation in transparent displays via the collected eye-tracking dataset (see Section~\ref{subsec:Data Collection}). The model performance is demonstrated in Fig.~\ref{fig:depth_model_results}. Results demonstrate loss and accuracy during training and evaluation (mean$\pm$STD). As shown, our proposed architecture can effectively approximate the gaze depth model in Eq.~\ref{eq:classical_gaze_depth_model}, achieving an average of $97.1\%$ test accuracy over five trials with different random seed initialization.

\textbf{Ablation Study} -- To confirm the effective integration of both the intra- and inter-stream self-attention layers, we conducted an ablation study by removing the intra-stream attention blocks (see Fig.~\ref{fig:gaze_depth_model}) and comparing the performance against the full architecture. Results are shown in Fig.~\ref{fig:depth_model_ablation_results} where as shown, the model performance drops significantly as compared to Fig.~\ref{fig:depth_model_results}, averaging around $65\%$ over five trials, after removing the intra-stream attention layers. This result confirms that the intra- and inter-stream self-attention layers can effectively capture the complex physical dependencies embedded in recorded eye tracking data.

\subsubsection{Gaze Depth Model Inference Results}
\label{subsubsec:Inference Results}
The operational feasibility and efficiency of the optimized multi-stream attention model for realtime inference was validated using Texas Instruments TDA4 VM series of System on Chip (SoC). TDA4 VM comprises of two ARM Cortex A-$72$ along with dedicated accelerators such as C7x Digital Signal Processor with Matrix Multiplication Accelerator. The hardware was chosen to represent a broad range of ARM based compute ubiquitous in most realtime, embedded devices. We evaluated the following key metrics: (1) accuracy, (2) model load time, (3) inference speed, and (4) total application time. Metrics (1) and (3) measure the ability of the model to accurately keep up with realtime data stream, while metrics (2) and (4) measure bring-up and tear-down time. Additionally, the metrics on the SoC were measured using two different frameworks: \textbf{PyTorch} and \textbf{ONNX Runtime} to account for discrepancies in software stacks. As a note, the ONNX Runtime is provided by the vendor, Texas Instruments, via a built-in package whereas the PyTorch implementation was ported on the board. Furthermore, the results were juxtaposed against metrics obtained by running similar experiments on a standard $x86$ laptop. The comparative analysis, conducted on a test subset of the collected dataset, is succinctly summarized in the Table~\ref{tab:inference_results}, illustrating our model's suitability for realtime deployment along our gaze tracking algorithm. In future, we can leverage advanced model quantization techniques~\cite{karimzadeh2020hardware1, karimzadeh2020hardware, karimzadeh2022bits, karimzadeh2022hardware} for an enhanced compression, achieving higher frequencies.
\begin{table}
\small
  \centering
  \begin{tabular}{@{}lcc@{}}
    \toprule
    Metric & ONNX Runtime & PyTorch \\
    \midrule
    Accuracy - \textbf{x86} & 98.38\% & 98.38\% \\
    Accuracy - \textbf{TDA4} & 98.38\% & 98.38\% \\
    \midrule
    \midrule
    Load Time - \textbf{x86} & 0.0061 [sec] & 0.0233 [sec] \\
    Load Time - \textbf{TDA4} & 0.0211 [sec] & 0.0309 [sec] \\
    \midrule
    \midrule
    Inference Speed - \textbf{x86} & 0.00007 [sec] & 0.0003 [sec] \\
    Inference Speed - \textbf{TDA4} & 0.0007 [sec] & 0.0151 [sec] \\
    \midrule
    \midrule
    Tot. Inference Time - \textbf{x86} & 0.0004 [sec] & 0.0019 [sec] \\
    Tot. Inference Time - \textbf{TDA4} & 0.0043 [sec] & 0.0906 [sec] \\
    \bottomrule
  \end{tabular}
  \caption{Inference evaluation for our gaze depth-level estimation model. Results show realtime feasibility of our model at inference.}
  \label{tab:inference_results}
  \vspace*{-0.5cm}
\end{table}


\section{Conclusion}
\label{sec:Conclusion}
We presented a systematic solution to realtime gaze target tracking and depth estimation from eye tracking data for Transparent Displays (TDs), e.g., dynamic Heads-Up Displays (HUDs) in vehicles. Our two-module solution includes: (1) a Quadtree-based gaze target tracking algorithm, and (2) a multi-stream attention model for estimating the depth-level of user gaze from eye tracking data, required to prevent undesired interactions with TDs. We collected real eye tracking data and demonstrated that not only our gaze monitoring system achieves promising accuracy in highly dynamic environments, but it also is realtime deployable. We presented various evaluation and ablation results, including inference performance on SoC evaluation boards. Our system sets a SOTA for human gaze monitoring on transparent AR displays with dynamic projected content. 

\bibliographystyle{unsrt}  
\bibliography{references}

\end{document}